\documentclass{emulateapj}
\shorttitle{Synchrotron Model of 3C 279 2015 June Flare}
\shortauthors{Asano \& Hayashida}
\begin{document}
\pagenumbering{arabic}
\title{
Synchrotron Gamma-Ray Emission Model of the Giant Outburst
of Quasar 3C 279 in 2015 June: Fast Reconnection or 
Stochastic Acceleration with Electromagnetic Cascade?
}
\author{Katsuaki Asano\altaffilmark{1}, and Masaaki Hayashida\altaffilmark{2,3}}
\email{asanok@icrr.u-tokyo.ac.jp, masaaki.hayashida@rikkyo.ac.jp}

\affil{\altaffilmark{1}Institute for Cosmic Ray Research, The University of Tokyo,
5-1-5 Kashiwanoha, Kashiwa, Chiba 277-8582, Japan}
\affil{\altaffilmark{2}Department of Physics, Rikkyo University, 3-34-1 Nishi-Ikebukuro,
Toshima-ku, Tokyo 171-8501, Japan}
\affil{\altaffilmark{3}Galaxies, Inc., 1-1-11 Minami-Ikebukuro, Toshima-ku, Tokyo, 171-0022, Japan}

\date{Submitted; accepted}

\begin{abstract}

We test the synchrotron emission scenario
for the very bright gamma-ray flare of blazar 3C 279 observed in 2015 June
using time-dependent numerical simulations.
A bulk Lorentz factor as high as 100 can bring the synchrotron maximum energy
above the GeV energy range.
We find two possible solutions for the X-ray to gamma-ray spectrum.
One is a prompt electron injection model with a hard power-law index
as magnetic reconnection models suggest.
A too strong magnetic field yields a too bright synchrotron X-ray flux
due to secondary electron--positron pairs.
Even in the prompt electron injection model,
the Poynting flux luminosity is at most comparable to
the gamma-ray or electron luminosity.
Another model is the stochastic acceleration model,
which leads to a very unique picture
accompanying the electromagnetic cascade and re-acceleration
of the secondary electron--positron pairs.
In this model,
the energy budget of the magnetic field is very low
compared to gamma rays and electrons.

\end{abstract}


\maketitle

\section{Introduction}
\label{sec:intro}

3C 279 is one of the most frequently studied flat spectrum radio quasars (FSRQs)
at redshift $z=0.536$.
The {\it Fermi}-Large Area Telescope (LAT) detected two prominent gamma-ray flares
in 2013 December and 2015 June \citep{hay15,3c279-16}.
The 2013 flare showed very hard spectrum with a photon index of $1.7 \pm 0.1$
above 100 MeV.
Leptonic scenarios for that flare imply extremely low magnetization
for the emission region \citep{hay15,asa15}.
In the 2015 flare, the gamma-ray flux is historically highest
with the gamma-ray isotropic luminosity of $\sim 10^{49}~\mbox{erg}~\mbox{s}^{-1}$.
The 2-minutes binned lightcurve shows a flux doubling timescale shorter than
5 minutes, which implies a very high bulk Lorentz factor such as $\Gamma>50$.

The standard model for the gamma-ray emission in FSRQs
is the inverse Compton scattering with external photons
from the broad line region (BLR) or dust torus
\citep[EIC model, e.g.][]{sik94}.
\citet{pet17} have discussed the 2015 June flare
adopting a proton synchrotron model.
Even with a super-Eddington jet luminosity and a smaller comoving source size
than $R/\Gamma$, where $R$ and $\Gamma$ are the distance from the central black hole
and the bulk Lorentz factor, respectively,
the electromagnetic cascade initiated by photomeson production
leads to a softer spectrum than the observed X-ray spectrum.
\citet{3c279-16} have proposed an interesting alternative scenario:
gamma rays are originating as electron synchrotron emission from
a highly magnetized plasma,
contrary to the case in the 2013 flare.

In this paper, we investigate possibilities of such leptonic synchrotron models
for the 2015 June gamma-ray flare with the time-dependent numerical
code in \citet{asa14}.
If the gamma-ray flare is attributed to synchrotron emission,
both the electrons emitting gamma rays and X-rays promptly lose their energies.
When electrons are injected with a power-law energy distribution of index $p>2$
in the energy range responsible for the photon emission from X-ray to gamma ray,
%with a index $p \geq 2$ following the standard shock acceleration model,
in the fast cooling regime the photon index becomes $(p+2)/2 \geq 2$ 
\citep[e.g.][]{der97}.
Yet, the observed X-ray photon index is $1.17 \pm 0.06$
\citep{3c279-16}, significantly harder than 2.
Even considering a minimum-energy of electrons at injection much higher
than the bulk Lorentz factor as assumed in gamma-ray bursts \citep{sar98},
%Alternatively, if the minimum energy of electrons at injection are much higher than
%the bulk Lorentz factor as assumed in gamma-ray bursts \citep{sar98},
the cooled electrons below the minimum-energy result in a photon index $1.5$.
Such a high minimum-energy at injection can be regarded as the extremely hard
limit for the low-energy portion of a broken power-law energy distribution.

In the synchrotron scenario,
acceleration mechanisms that produce a hard spectrum of electrons are required.
In this paper, we adopt a prompt power-law injection model
($p<2$) and a stochastic acceleration model,
which are motivated by magnetic reconnection and
turbulence acceleration, respectively.

\section{Model Setup}
\label{sec:model}

In this paper, we adopt
the time-dependent numerical code in \citet{asa14}
\citep[see also,][]{asa15,asa18}.
In this code, the geometry of the jet is conical,
and the evolution of the electron/photon energy distribution
is calculated taking into account electron injection,
photon production via synchrotron and inverse Compton,
photon escape, radiative and adiabatic cooling of electrons,
synchrotron self-absorption, $\gamma \gamma$ pair production,
and attenuation by the extragalactic
background light (EBL).
Our code is based on the one-zone approximation
so that the secondary pairs are injected into the same
region and experience the same magnetic field.
The 3C 279 photon spectrum shown in \citet{3c279-16}
was obtained by averaging the flux over the orbital period of $95.6$ minutes,
which is much longer than the variability timescale.
We took the photon spectra of the highest-flux orbit (Orbit-C) and the
subsequent orbit (Orbit-D) for our studies of the emission modeling.
The minute-scale variability in flux was observed in both the orbits,
and the simultaneous {\it Swift} observational data for X-ray and UV ({\it W}2)
bands are available during Orbit D.
In this paper, we obtain steady solutions of the emission,
which can be regarded as the average emission over the orbital period.
In this treatment, the temporal evolution of the electron/photon energy distribution
in the jet frame is equivalent to the radial evolution in the steady jet
\citep[see][for details]{asa14}.

The jet parameters are the initial radius $R_0$,
the bulk Lorentz factor $\Gamma$, and the initial magnetic field $B_0$.
The jet opening angle is assumed as $\theta_{\rm j}=1/\Gamma$. 
The emission released from various angles within $\theta_{\rm j}$ is integrated
with the exact beaming factor to obtain a spectrum for an on-axis observer.
Though we cannot uniquely determine the parameter values,
throughout this paper (except for model C0),
we adopt $R_0=7.1\times 10^{16}$ cm
and $\Gamma=100$, yielding
a variability timescale $t_{\rm var} \simeq R_0/(c \Gamma^2)=237$ s
consistent with the observed one.
Adjusting other parameters concerning electron injection/acceleration,
we try to reproduce the gamma-ray flux by synchrotron emission.
The adopted value of $\Gamma$ is larger than the typical value
for blazars ($\Gamma \sim 10$), but not unheard of.
For example, the 2006 July flare of PKS 2155--304,
which showed a gamma-ray ($>200$GeV) variability on timescales of $\sim 200$ s,
requires a large Lorentz factor $\gtrsim 100$ to reconcile
the variability with the broadband spectrum \citep{aha07,kus08}.
As will be shown below, such a large $\Gamma$ is required to
explain the spectrum from X-ray to gamma ray by synchrotron emission.
%Also for the extreme flare in 3C 279, such a large $\Gamma$
%would be applicable.

We divide the conical jet into shells with width of $R_0/\Gamma^2$
in the observer frame.
The comoving volume of each shell for the one-side jet evolves as
%%%%%%%%%%%%%%%%%%%%%%%
\begin{equation}
V'_{\rm j}=\frac{4 \pi R_0^3}{\Gamma} \left(\frac{R}{R_0} \right)^2
\frac{(1-\cos{\theta_{\rm j}})}{2},
\end{equation}
%%%%%%%%%%%%%%%%%%%%%%%
with a distance $R$.
Considering continuous ejection of identical shells from $R=R_0$,
a steady outflow is realized in our computation.
The magnetic field
in the jet frame evolves as
%%%%%%%%%%%%%%%%%%%%%%%
\begin{eqnarray}
B'=B_0 \left( \frac{R}{R_0} \right)^{-1}.
\end{eqnarray}
%%%%%%%%%%%%%%%%%%%%%%%

The jet is surrounded by an external photon field coming from the BLR.
The energy density and spectrum of the external photon filed in the jet frame
are provided by the same model in \citet{hay12};
the photon spectrum is the diluted Planck distribution
with photon temperature of
%%%%%%%%%%%%%%%%%%%%%%%
\begin{eqnarray}
T'_{\rm UV}=10 \Gamma~\mbox{eV},
\end{eqnarray}
%%%%%%%%%%%%%%%%%%%%%%%
and the energy density is written as
%%%%%%%%%%%%%%%%%%%%%%%
\begin{eqnarray}
U'_{\rm UV}=\frac{0.1 \Gamma^2 L_{\rm D}}{3 \pi c R^2_{\rm BLR} (1+(R/R_{\rm BLR})^3)},
\end{eqnarray}
%%%%%%%%%%%%%%%%%%%%%%%
\citep{sik09},where $L_{\rm D}$ is the disk luminosity, and
$R_{\rm BLR}$ is the size of BLR,
%%%%%%%%%%%%%%%%%%%%%%%
\begin{eqnarray}
R_{\rm BLR}=0.1 \left( \frac{L_{\rm D}}{10^{46}~\mbox{erg}~\mbox{s}^{-1}} \right)^{1/2}~\mbox{pc}.
\end{eqnarray}
%%%%%%%%%%%%%%%%%%%%%%%
For 3C 279, we adopt $L_{\rm D}=2 \times 10^{45}~\mbox{erg}~\mbox{s}^{-1}$ \citep{pia99}.
The equipartition magnetic field with the photon energy density is
$B_{\rm eq}=91$ G at $R=R_0$.
In \citet{3c279-16}, the equi-partition strength was estimated
as $1.3$ kG because of a lower bulk Lorentz factor
($\Gamma=25$) they assumed ($B_{\rm eq} \propto \Gamma^{-3}$).
Note that the high photon temperature in the comoving frame
makes the Klein--Nishina effect significant for electrons
with Lorentz factor $\gamma'_{\rm e}>m_{\rm e} c^2/(4 T'_{\rm UV}) \simeq 130$.
Even for a lower magnetic field than $B_{\rm eq}$,
the electron cooling is
sensitive to the value of $B'$.

\section{Prompt Power-Law Injection}
\label{sec:PL}

\begin{figure*}[!t]
\centering
\epsscale{1.1}
\plottwo{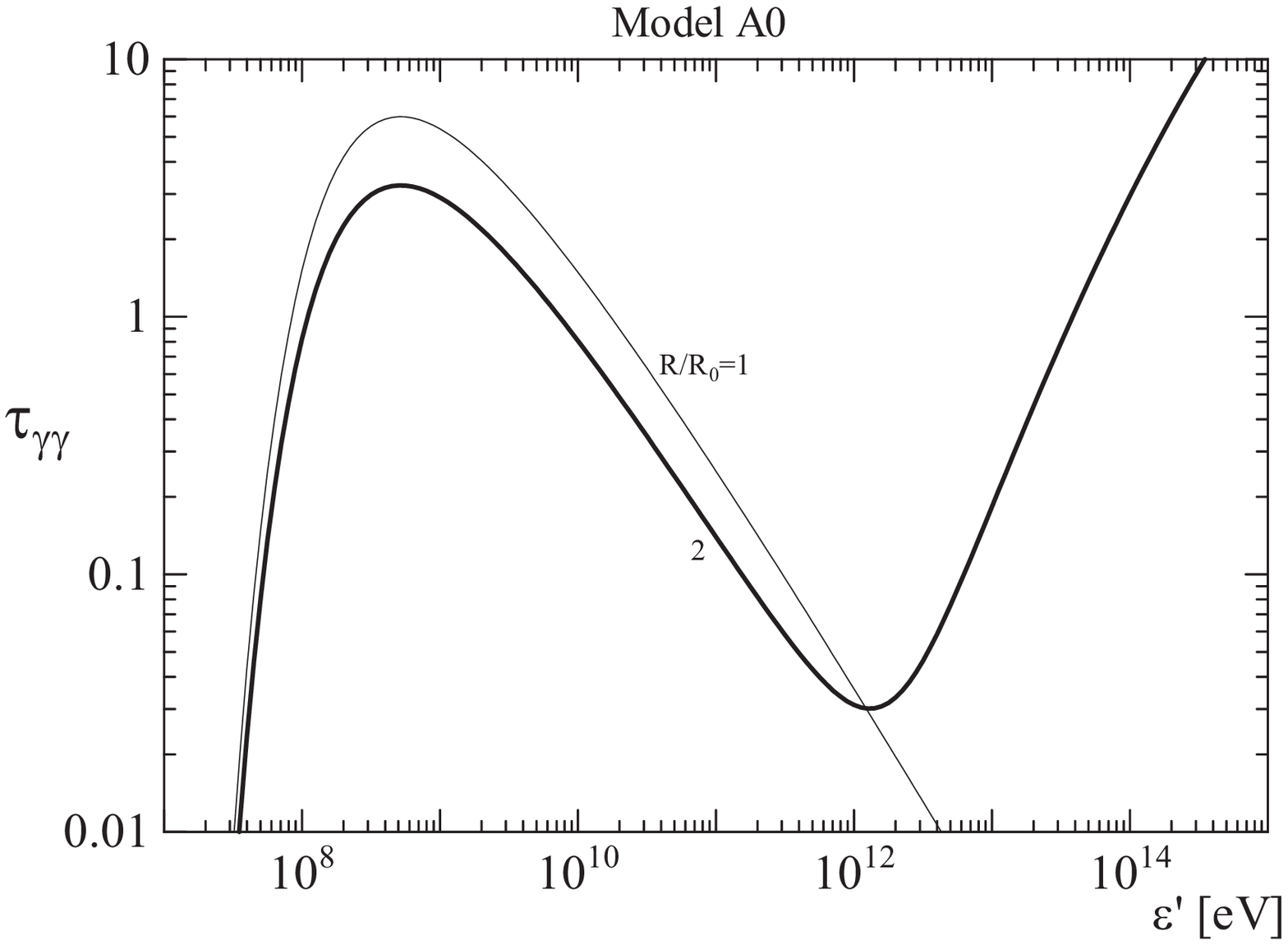}{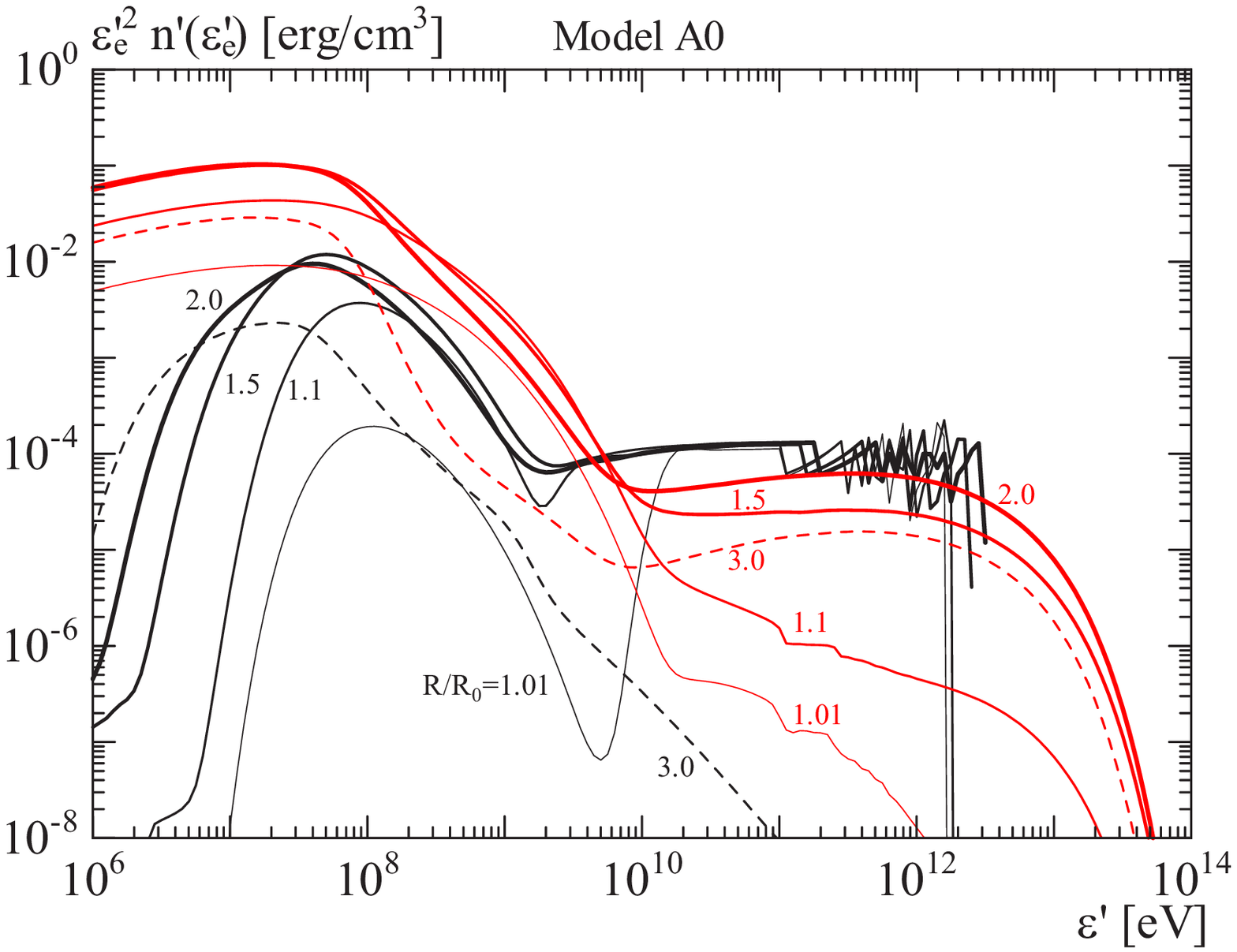}
\caption{(Left) Optical depth for $\gamma \gamma$-absorption
at $R=R_0$ (thin) and $2 R_0$ (thick) in model A0. The photon energy is measured
in the jet comoving frame.
(Right) Evolution of the electron (black) and photon (red) energy distributions
with increasing distance $R$ (thin to thick)
in model A0. The numbers beside each line denote
$R/R_0$. Electrons are injected between $R=R_0$ and $2R_0$.
At $R=3R_0$ (thin dashed lines), the electron injection had been already stopped.
\label{fig:fadd}}
\end{figure*}

\begin{figure}[!t]
\centering
\epsscale{1.1}
\plotone{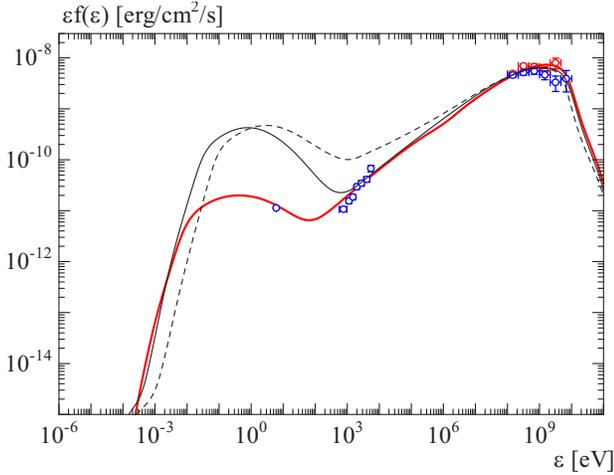}
\caption{Photon spectra of the prompt power-law injection models.
The model parameters are in Table \ref{table:para0}.
The red solid, thin black solid, and black thin dashed lines show
the spectra for model A0, B0,
and C0, respectively.
The red and blue data points are observed data
for Orbits C and D, respectively, in \citet{3c279-16} and \citet{hay17}.
\label{fig:f0}}
\end{figure}

First, motivated by the magnetic reconnection model,
we test cases for electron injection with a power-law energy distribution
with an exponential cut-off:
$\dot{n}'(\gamma'_{\rm e}) \propto \gamma'^{-p}_{\rm e}
\exp{(-\gamma'_{\rm e}/\gamma_{\rm max})}$.
The strong magnetic field in our synchrotron model leads to
prompt cooling of electrons.
If the minimum electron energy at injection is low enough,
the hard X-ray spectrum (see Figure \ref{fig:f0})
is inconsistent with the fiducial index $p \sim 2$
in the standard shock acceleration.
However, the electron acceleration by magnetic reconnection can
produce a very hard spectrum \citep[e.g.,][and references therein]{sir14}.
An electron injection with $p<2$ is almost equivalent
to a monoenergetic injection at the maximum energy.
In this case, the prompt cooling
yields a hard synchrotron spectrum with a photon index of 1.5.

\begin{table}[!htb]
  \caption{Prompt Power-Law Injection Model: Parameters}
  \label{table:para0}
  \centering
  \begin{tabular}{lllllll}
\hline
    Model & $\Gamma$ & $R_0$ & $B_0$ &  $p$  & $\gamma_{\rm max}$ & $L_{\rm e,inj}$  \\
          &          & [$10^{16}$~cm] & [G] & & & [$\mbox{erg}~\mbox{s}^{-1}$] \\
    \hline \hline
    A0	& 100 & $7.1$ & 8.0 & $0.0$ & $1.6 \times 10^{7}$ & $2.7 \times 10^{45}$ \\
    B0	& 100 & $7.1$ & 80.0 & $0.2$ & $5.0 \times 10^{6}$ & $2.7 \times 10^{45}$  \\
    C0	& 30 & $0.65$ & 110.0 & $0.2$ & $7.7 \times 10^{6}$ & $2.7 \times 10^{45}$  \\
    \hline
  \end{tabular}
\tablecomments{The electron injection luminosity $L_{\rm e,inj}$ includes that for the counter jet.
}
\end{table}

Table \ref{table:para0} shows the model parameters.
We inject electrons from the initial radius $R_0$
in the dynamical timescale $R_0/c$ with
luminosity of $L_{\rm e,inj}$
and very hard spectral index.
Since almost all energy injected as electrons is converted
to gamma-ray emission, the injection luminosity $L_{\rm e,inj}$
is the same for all models A0, B0, and C0.
The minimum Lorentz factor $\gamma_{\rm min}=100 \ll \gamma_{\rm max}$
is not an important parameter; the injected energy at $\gamma'_{\rm e} \sim \gamma_{\rm min}$
is negligible.
Model B0 has a stronger magnetic field but lower maximum energy $\gamma_{\rm max}$
than those in model A0.
To make a flat spectrum in the GeV energy range by the curved electron energy distribution
around $\gamma_{\rm max}$, we need to keep the typical synhcrotron
photon energy $\propto \Gamma B_0 \gamma_{\rm max}^2$
constant.
In all the models, the combination $\Gamma B_0 \gamma_{\rm max}^2$
is adjusted to the same value, so that the gamma-ray spectra
are almost identical in spite of different magnetic fields.

When we write the acceleration timescale as $t_{\rm acc}=\xi \gamma_{\rm e} m_{\rm e}c/(eB)$,
where $\xi$ is a dimensionless parameter,
the balance with the cooling timescale, $t_{\rm c,syn}=
6 \pi m_{\rm e} c/(\sigma_{\rm T} B^2 \gamma_{\rm e})$,
provides us
the maximum energy
%%%%%%%%%%%%%%%%%%%%%%%
\begin{eqnarray}
\gamma_{\rm max}\simeq \sqrt{\frac{6 \pi e}{\xi \sigma_{\rm T} B}}
=3.7 \times 10^7 \xi^{-1/2} \left( \frac{B}{10~\mbox{G}} \right)^{-1/2}.
\label{eq:max}
\end{eqnarray}
%%%%%%%%%%%%%%%%%%%%%%%
Usually, $\xi$ is assumed larger than unity,
but the particle acceleration by magnetic reconnection
would attain $\xi<1$ \citep{cer13}.
The values of $\gamma_{\rm max}$ in Table \ref{table:para0}
imply that the acceleration efficiency is close to the limit of $\xi=1$.
The maximum synchrotron photon energy is independent of the magnetic field
as
%%%%%%%%%%%%%%%%%%%%%%%
\begin{eqnarray}
\varepsilon_{\rm max} &\simeq& \Gamma \frac{3}{2}\frac{\hbar e B'}{m_{\rm e} c} \gamma'^{2}_{\rm max}
=24 \xi^{-1} \left( \frac{\Gamma}{100} \right)~\mbox{GeV}.
\label{eq:phmax}
\end{eqnarray}
%%%%%%%%%%%%%%%%%%%%%%%

The dominant target photons for $\gamma \gamma$-absorption
is the external photon field.
As the left panel in Figure \ref{fig:fadd} shows,
the absorption effect becomes significant above
$\varepsilon'_{\rm cut}\sim(m_{\rm e} c^2)^2/T'_{\rm UV}
\sim 10^8$ eV in the comoving frame.
The cut-off energy for an observer is expected as $\Gamma \varepsilon'_{\rm cut} \sim
10$ GeV, which is almost independent of $\Gamma$.

In our code, while the calculation time step, which depends on electron energy,
is always much shorter than the electron cooling timescale,
the time steps for electron injection and calculation output are
longer than the cooling timescale in the high energy range.
Because of this time step effect,
the electron energy distribution above $10^{11}$ eV is noisy
in the right panel in Figure \ref{fig:fadd}.
The electrons above $10^{12}$ eV have been already cooled
for this output time step.
Since the emission is integrated with significantly short time steps,
the photon spectrum is not affected by the output time step.
As the electron spectra in Figure \ref{fig:fadd} shows,
the cooled electrons distribute
$n(\gamma'_{\rm e}) \propto \gamma'^{-2}_{\rm e}$
below $\gamma_{\rm max}$, which yields
a photon spectrum $f(\varepsilon) \propto \varepsilon^{-0.5}$.
However, the observed X-ray index is $0.2$.
To explain this harder X-ray spectrum, our numerical
model needs additional parameters that describe a more complicated process/situation
such as particle escape, and the decay or the inhomogeneity of magnetic field.
We can see the spectral bump due to secondary electron--positron pairs
around $10^8$ eV in the electron spectra.

\begin{figure*}[!t]
\centering
\epsscale{1.1}
\plottwo{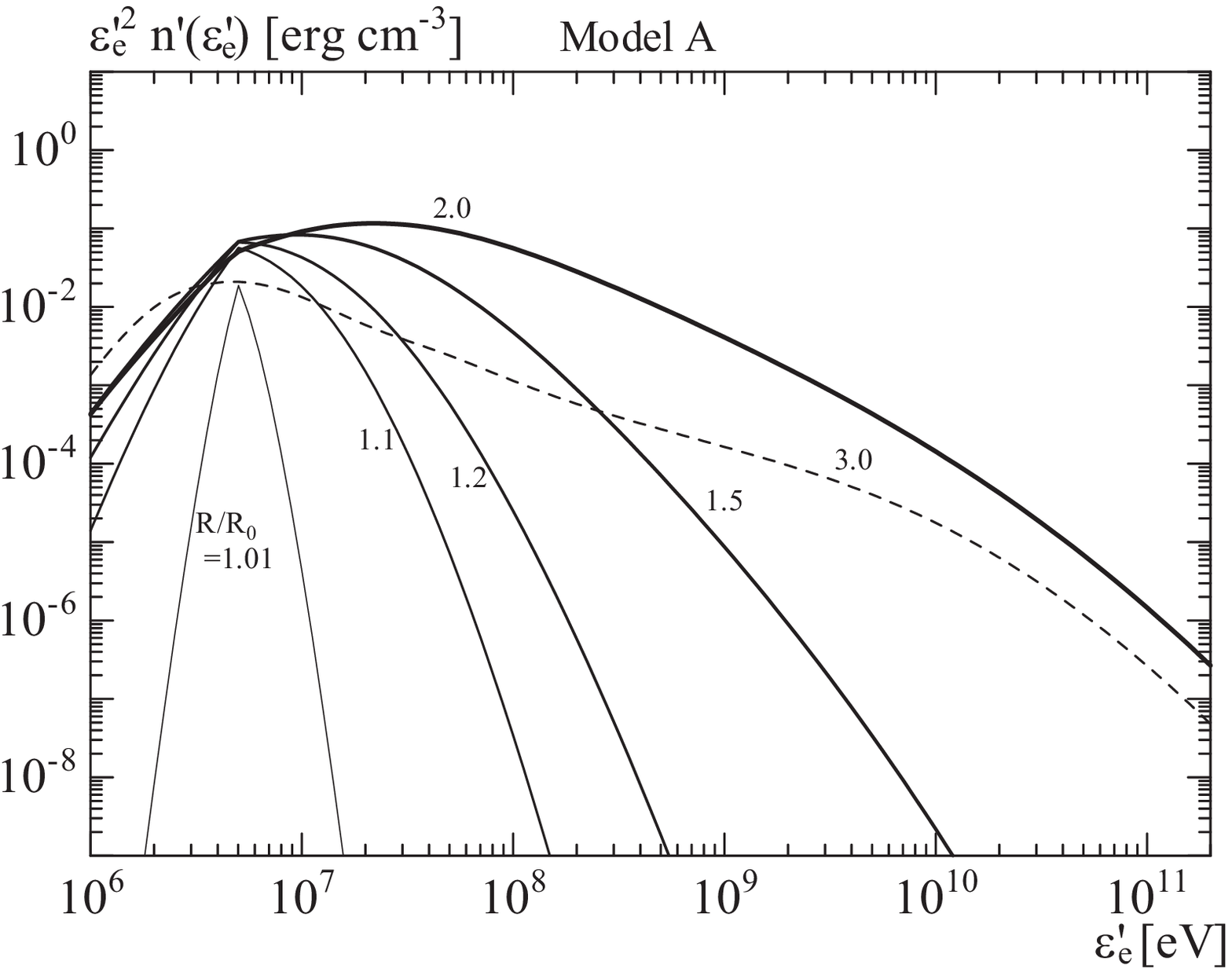}{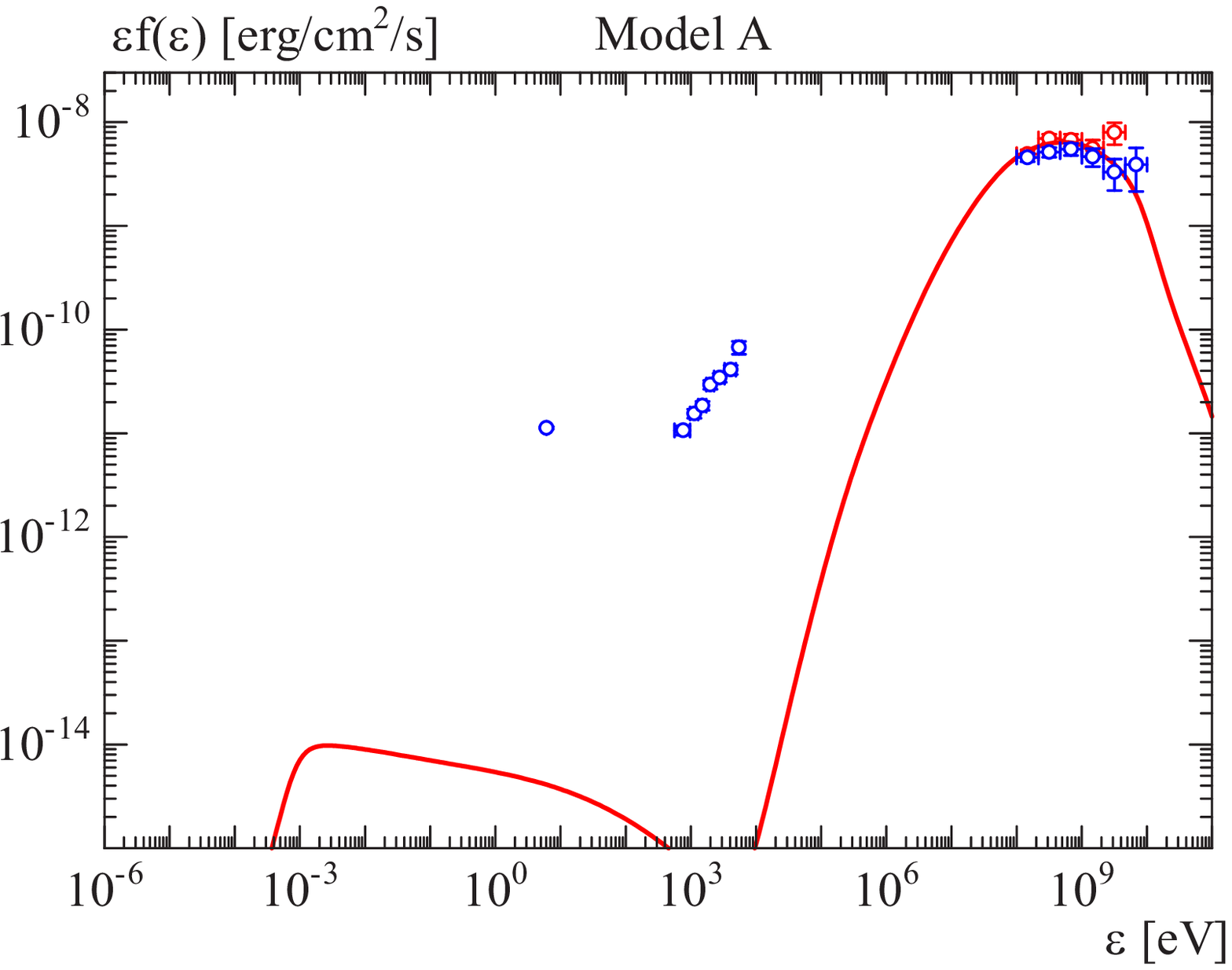}
\caption{(Left) Evolution of the electron energy distribution
(including secondary electron--positron pairs)
with increasing distance $R$ (thin to thick)
in model A. The numbers beside each line denote
$R/R_0$.
Electrons are injected and accelerated between $R=R_0$ and $2R_0$.
At $R=3R_0$ (thin dashed line), the electron injection
and acceleration had been already stopped.
(Right) Photon spectrum for model A (red).
The observed data are the same as in Fig. \ref{fig:f0}.
\label{fig:A}}
\end{figure*}

Model A0 in Figure \ref{fig:f0} seems consistent with
the gamma-ray spectrum and the UV flux.
The flux level of the X-ray emission is also reproduced,
though the spectrum is softer
than the observed data as we have mentioned.
The steep cut-off above 10 GeV is due to electron--positron
pair absorption.
The bump below 100 eV is the synchrotron emission
by secondary electron--positron pairs.
The increase of the optical depth
above $1$ TeV in Figure \ref{fig:fadd} at $R=2R_0$
is due to the growth of the target photons for $\gamma \gamma$-absorption
in the optical/IR band by this secondary synchrotron emission.
In model B0, the higher secondary bump disagrees with the observed X-ray
and UV flux. 
The magnetic luminosity, including the contribution of the counter jet,
is calculated as
%%%%%%%%%%%%%%%%%%%%%%%
\begin{eqnarray}
L_{B}&=&4 \pi R_0^2 \Gamma^2 \left( \frac{B_0^2}{8 \pi} \right) c (1-\cos{\theta_{\rm j}}) 
\simeq \frac{1}{4} R_0^2 B_0^2 c \\
&\simeq&  2.4 \times 10^{45}
\left( \frac{B_0}{8~\mbox{G}} \right)^2 \left( \frac{R_0}{7.1 \times 10^{16}~\mbox{cm}} \right)^2
\mbox{erg}~\mbox{s}^{-1}, \nonumber \\
\label{eq:Lb}
\end{eqnarray}
%%%%%%%%%%%%%%%%%%%%%%%
which is comparable to the electron luminosity in model A0.
Model B0 has difficulty also in the context of the energy budget,
because the total magnetic luminosity of the source exceeds the
Eddington luminosity;
the Eddington luminosity is
$8 \times 10^{46}\mbox{erg}~\mbox{s}^{-1}$ for
the black hole mass of $5 \times 10^8 M_\odot$.
Therefore, the largely Poynting-flux-dominated jet,
like model B0, is unlikely.

Since the observed variability provides an upper limit for $R_0/\Gamma^2$,
a similar model to model A0 with a shorter $R_0$ keeping $B_0=8$ G
and $L_{\rm e,inj}$
also yields the same successful spectrum.
However, according to equation (\ref{eq:Lb}),
such a model leads to $L_{\rm e,inj} \gg L_{B}$,
which does not seem preferable for the magnetic reconnection model.

In model C0, we test a case with a lower Lorentz factor maintaining
the variability timescale $R_0/(c \Gamma^2)$,
typical synchrotron photon energy $\propto \Gamma B_0 \gamma_{\rm max}^2$
and the luminosities $L_{\rm e,inj}$ and $L_{B}$.
The smaller $R_0$ required by the small $\Gamma$
leads to a large magnetic field $B_0 \simeq 110$ G.
Therefore, the secondary synchrotron emission becomes too luminous
(see the thin dashed line in Figure \ref{fig:f0}).
A high Lorentz factor is required for the prompt power-law injection model
with the condition $L_{B} \gtrsim L_{\rm e,inj}$.

\section{Stochastic Acceleration Model}
\label{sec:SSA}

Here, let us consider the case in which electrons
are gradually accelerated by turbulence
as discussed in \citet{asa14,asa15,asa18}.
The turbulence acceleration process,
equivalent to the second order Fermi acceleration,
can produce a harder spectrum than the standard shock acceleration.
In this section, we show and compare results for both
the EIC (models A and B) and the synchrotron (model C) emission scenario
for gamma rays.
As will be shown below, the EIC models have difficulty in reconciling
the gamma-ray with the X-ray and optical data.

The turbulence acceleration
is expressed by the energy diffusion coefficient,
%%%%%%%%%%%%%%%%%%%%%%%
\begin{equation}
D'_{\gamma \gamma}=K \gamma'^2_{\rm e},
\label{difdif}
\end{equation}
%%%%%%%%%%%%%%%%%%%%%%%
where the parameter $K$ is constant.
In this paper, we assume the hard-sphere type
acceleration ($D'_{\gamma \gamma} \propto \gamma'^2_{\rm e}$).
As \citet{asa15,asa18} show, the hard-sphere acceleration
can reproduce broadband spectra for several blazars.
The acceleration timescale ($\sim K^{-1}$)
in this model is constant
irrespective of particle energy.
While the acceleration timescale for the highest-energy particles
can be comparable to that frequently assumed in the shock acceleration models,
lower-energy particles are accelerated with much longer timescales
than their gyromotion period.
The hard-sphere type acceleration can be caused by
large-scale hydrodynamical eddy-turbulence
\citep[e.g.][]{pt88,cl06} or compressional magnetohydrodynamic waves
\citep{ter19}.
When the magnetic field energy is dominant (model B),
the turbulence acceleration would show a different behavior
from that in a weakly magnetized plasma
\citep{dem19}.
However, for simplicity, we use equation (\ref{difdif}) 
throughout this section.

\begin{figure*}[!t]
\centering
\epsscale{1.1}
\plottwo{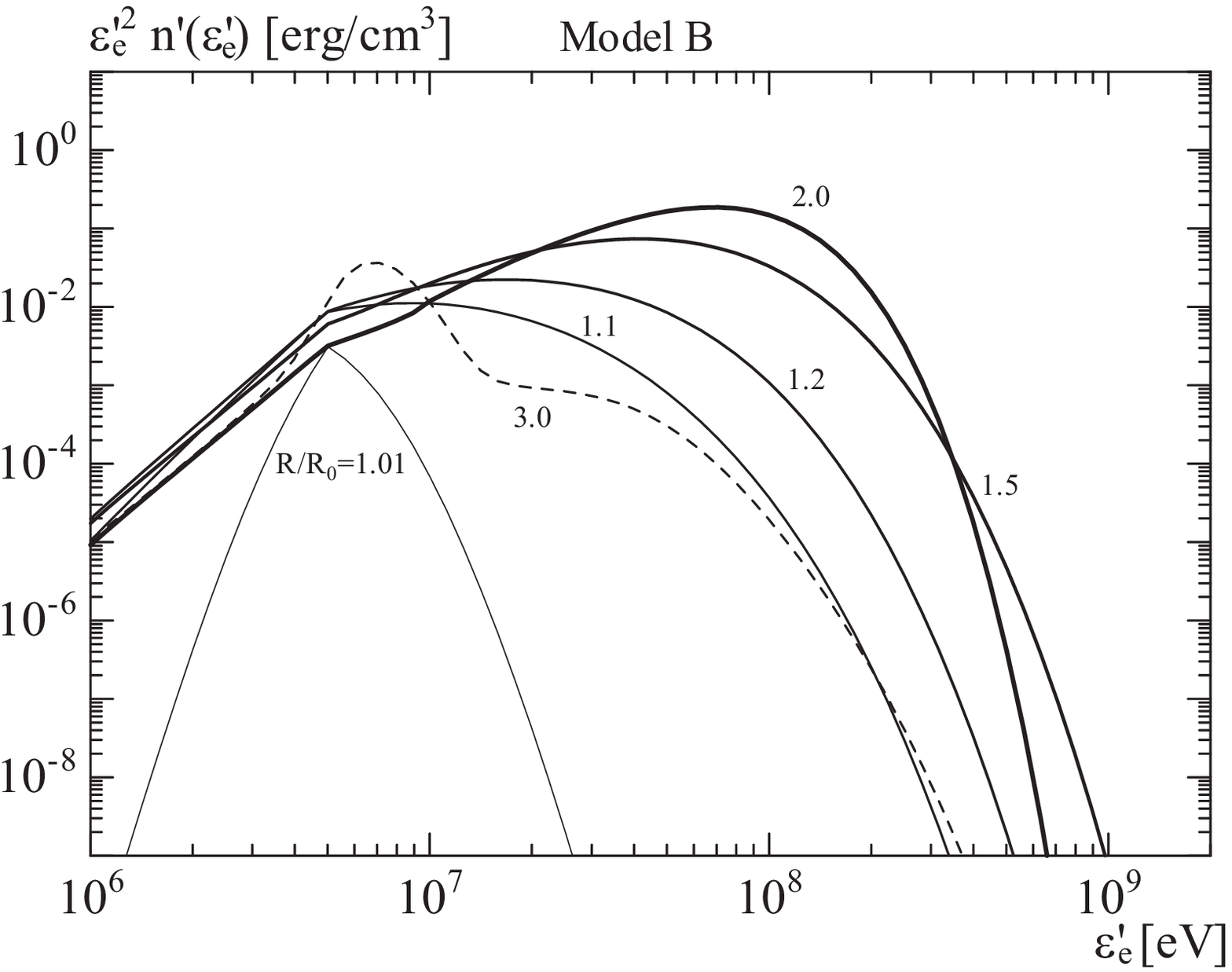}{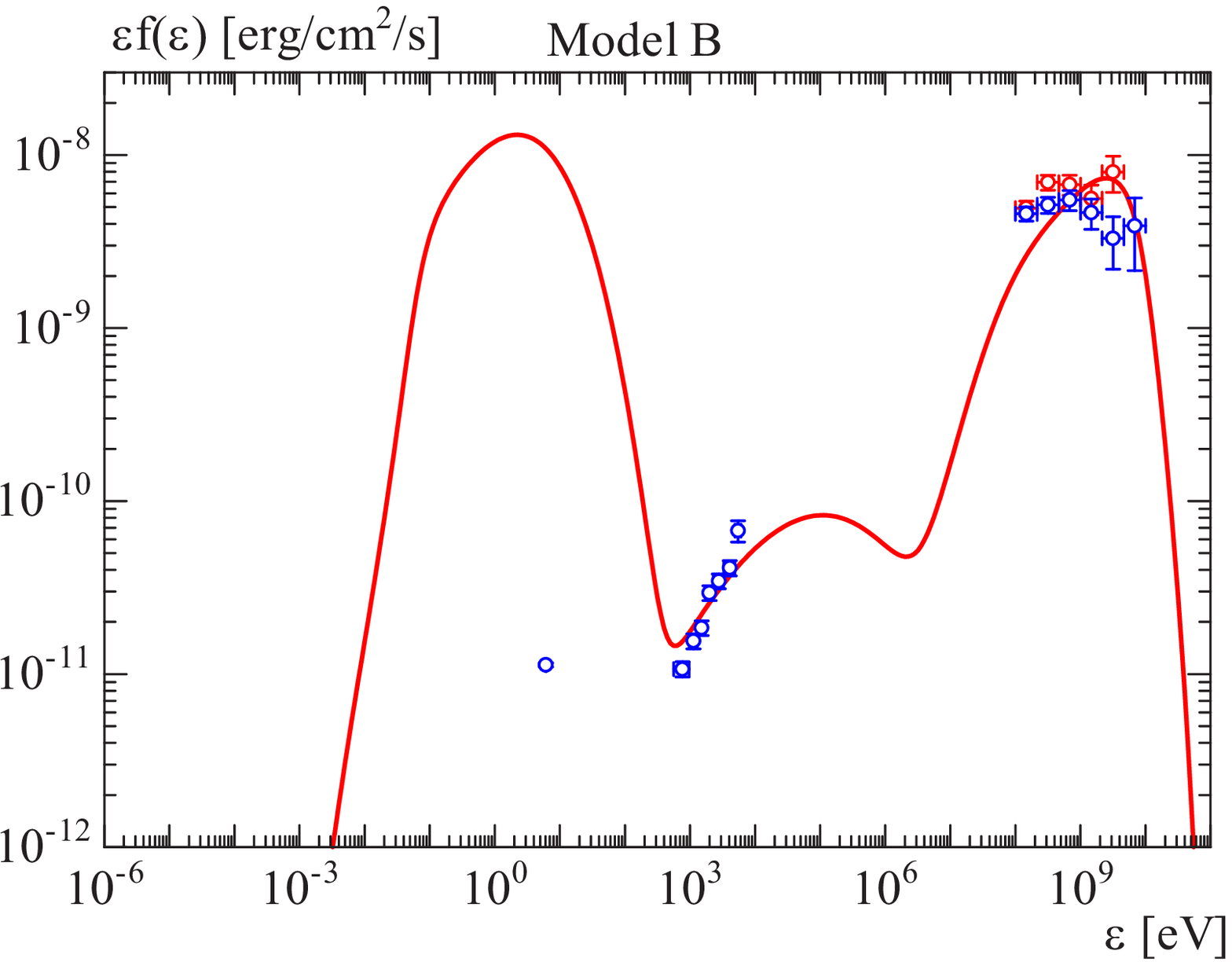}
\caption{(Left) Evolution of the electron energy distribution
(including secondary electron--positron pairs)
with increasing distance $R$ (thin to thick)
in model B. The numbers beside each line denote
$R/R_0$.
Electrons are injected and accelerated between $R=R_0$ and $2R_0$.
At $R=3R_0$ (thin dashed line), the electron injection
and acceleration had been already stopped.
(Right) Photon spectrum for model B (red).
The observed data are the same as in Fig. \ref{fig:f0}.
\label{fig:B}}
\end{figure*}

We inject electrons at a constant rate $\dot{N}_{\rm j}$
into the volume $V'_{\rm j}$ from $R=R_0$ to $2 R_0$.
The initial Lorentz factor of electrons is $\gamma'_{\rm inj}=10$.
For $R > 2 R_0$, the injection and energy diffusion are halted.

\begin{table}[!htb]
  \caption{Stochastic Acceleration Model: Parameters and Energy Density Ratio}
  \label{table:para1}
  \centering
  \begin{tabular}{lllll}
\hline
    Model & $B_0$ [G] &  $K$ [$\mbox{s}^{-1}$] & $\dot{N}_{\rm j}$ [$\mbox{s}^{-1}$]
    & $U_B/U_{\rm e}$\footnote{The energy density ratio at $R=2 R_0$.} \\
    \hline \hline
    A	 & 0.1 & $4.2 \times 10^{-5}$ & $2.5 \times 10^{45}$ & $2.7 \times 10^{-4}$ \\
    B	 & 80 & $1.3 \times 10^{-4}$ & $6.7 \times 10^{44}$ & $180$  \\
    C & 0.1 & $4.2 \times 10^{-4}$ & $4.9 \times 10^{32}$ & $1.2 \times 10^{-4}$  \\
    \hline
  \end{tabular}
\tablecomments{The other parameters $\Gamma=100$
and $R_0=7.1 \times 10^{16}$ cm are common.
}
\end{table}

The model parameters are summarized in Table \ref{table:para1}.
%A constraint for synchrotron emission
%in UV band for 2015 June flare is the UV data point
%as shown in Figures \ref{fig:f0}--\ref{fig:C}.
The shape of the spectrum is determined by the combination of $B_0$
and $K$. The injection rate $\dot{N}_{\rm j}$
is adjusted to match the gamma-ray flux level.
The UV flux may have a different origin from that for the gamma-ray flare,
as discussed in \citet{asa15}. The UV data point
can be regarded as the upper limit for the synchrotron component.
First, in model A, we test the usual external inverse Compton (EIC) model.
The flat gamma-ray spectrum is reproduced by Compton-scattered
UV photons as shown in Figure \ref{fig:A}.
However, the narrow EIC spectral hump does not agree with the observed X-ray flux.
Another component such as emission from a different region
is needed for X-ray emission in this case.
In our model, as electrons are injected and accelerated as far as $R=2R_0$,
the electron energy density becomes maximum at $R=2R_0$.
The energy density ratio of the magnetic field to the electron energy density
$U_B/U_{\rm e}$ at $R=2R_0$ is listed in Table \ref{table:para1}.
In model A,
the magnetic field is much weaker than the equipartition value.

In model B, we adopt
a stronger magnetic field.
As shown in Figure \ref{fig:B},
in spite of the larger diffusion coefficient,
the cooling effect by the strong magnetic field suppresses
the maximum energy of electrons compared to model A.
Model B roughly reproduces the X-ray spectrum
by synchrotron self-Compton emission in addition to
the EIC gamma-ray spectrum.
However, the synchrotron flux in the optical-UV band is
extremely brighter than the UV data point and
historical data by three orders of magnitude.
Furthermore, as mentioned for model B0 in the previous section,
$B_0=80$ G seems too large compared to the Eddington luminosity.

\begin{figure*}[!t]
\centering
\epsscale{1.1}
\plottwo{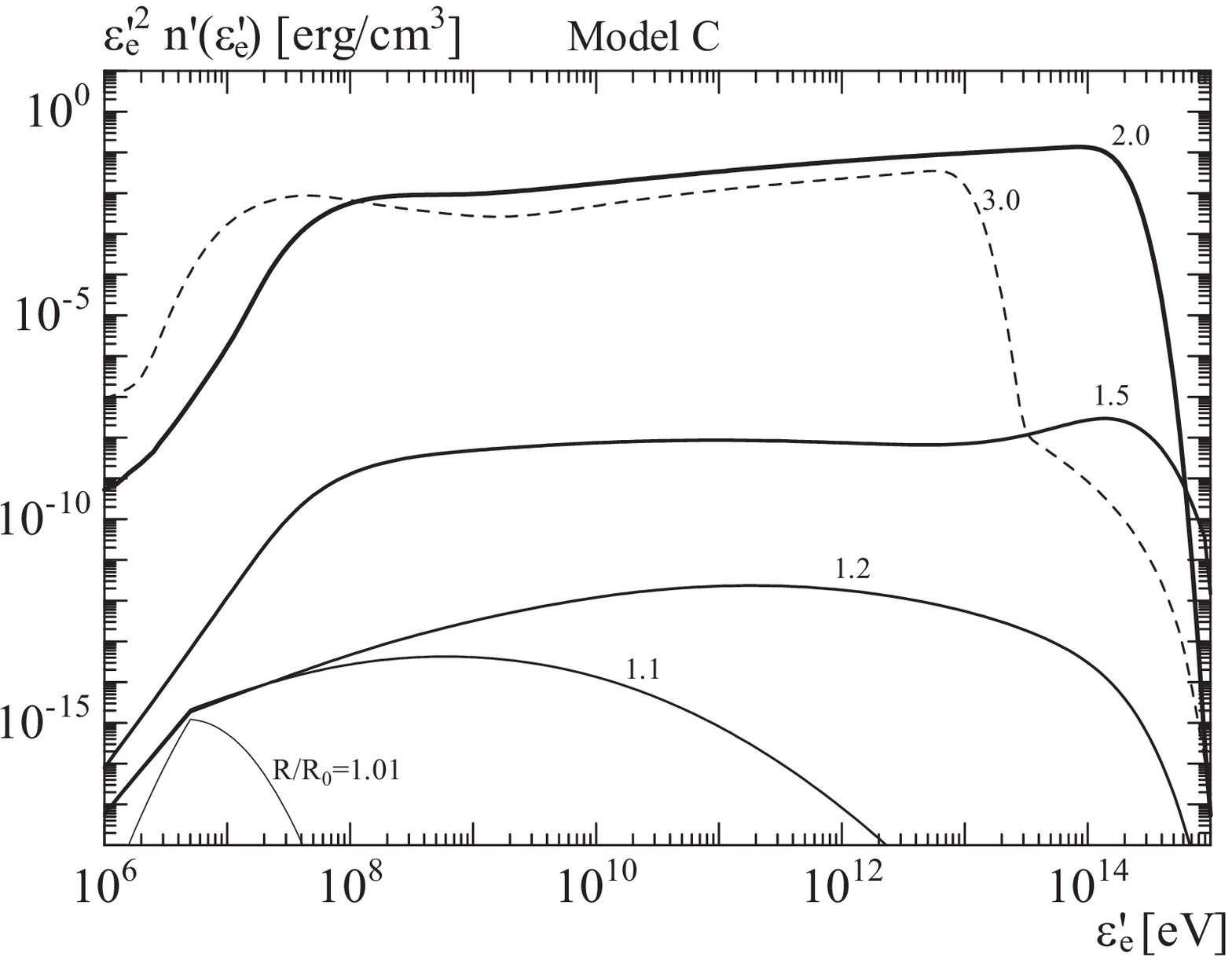}{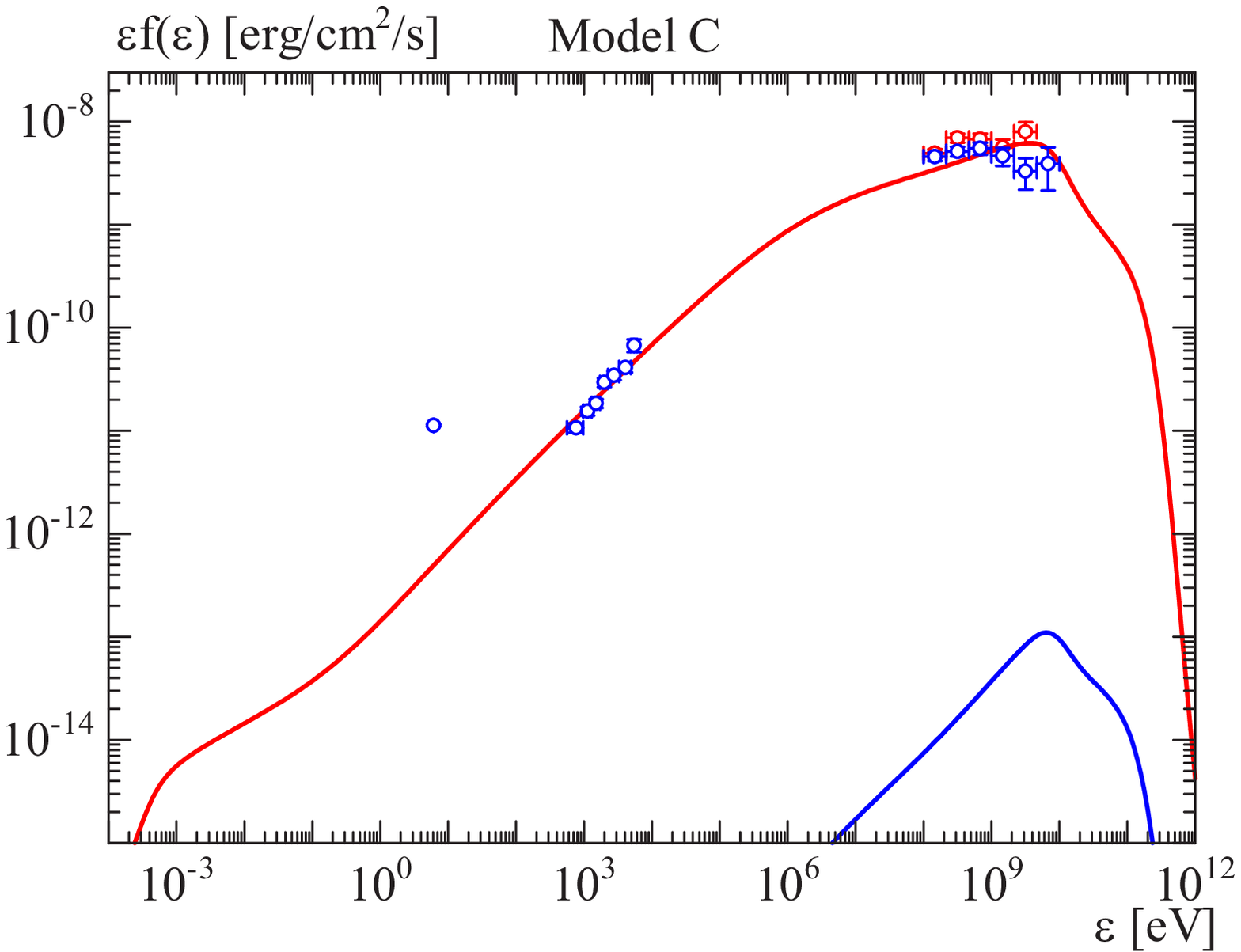}
\caption{(Left) Evolution of the electron energy distribution
(including secondary electron--positron pairs)
with increasing distance $R$ (thin to thick)
in model C. The numbers beside each line denote
$R/R_0$.
Electrons are injected and accelerated between $R=R_0$ and $2R_0$.
At $R=3R_0$ (thin dashed line), the electron injection
and acceleration had been already stopped.
(Right) Photon spectrum for model C (red).
The blue line shows the case neglecting the secondary
electron--positron pairs.
The observed data are the same as in Fig. \ref{fig:f0}.
\label{fig:C}}
\end{figure*}

As the gamma-ray photon density is constrained by the observed flux,
the numbers of the secondary pairs in both models A and B
would be the same order as those in models A0 or B0.
The low magnetic field in model A and the high synchrotron flux by the primary
electrons in model B inhibit emergence of a distinct spectral component due to secondary pairs.

In our stochastic acceleration model, to increase the maximum electron energy,
a lower magnetic field is preferable.
In model C, we decrease the magnetic field again, and increase
the diffusion coefficient.
The cascade process with re-acceleration, which will be discussed later,
produces a flat electron/positron energy distribution, as shown in
Figure \ref{fig:C}.
As we have mentioned in section \ref{sec:model},
electrons of $\gamma'_{\rm e}>130$
($\varepsilon'_{\rm e}=\gamma'_{\rm e} m_{\rm e} c^2 \sim 70$ MeV) are in the
Klein--Nishina regime for the target photons of $4 T'_{\rm UV}=4$ keV.
The typical photon energy of IC emission in this regime
is $\sim \Gamma \gamma'_{\rm e} m_{\rm e} c^2$.
The electron energy distribution
and the EIC photon spectrum of the previous EIC model, A (Figure \ref{fig:A}),
show that electrons around $
\varepsilon'_{\rm e}=70$ MeV emit GeV gamma rays via EIC.
In model C, electrons/positrons with much higher energies than $\varepsilon'_{\rm e}=70$ MeV
energetically dominate. Although such high-energy particles can emit gamma rays via EIC emission
in spite of the Klein--Nishina effect, the EIC photon energy
($\varepsilon \gg 70 \Gamma$ MeV $=7$ GeV)
is much higher than the energy range of {\it Fermi}.
The IC photons emitted by electrons with $\varepsilon'_{\rm e} \gg 70$ MeV
are promptly absorbed,
and produce secondary electron--positron pairs.
As a result, the IC photons emitted by the high-energy particles almost do not contribute
to the flux below 7 GeV.
Only gamma rays in a narrow energy range between $\varepsilon'_{\rm cut} \sim 100$ MeV
and $2 \times 70$ MeV produce secondary pairs that emit gamma rays below 7 GeV
via EIC.

Even though the magnetic energy density in model C is much lower
than the photon energy density, high-energy electrons/positrons
emit synchrotron photons, and cool mainly via synchrotron because of the Klein--Nishina effect.
The maximum electron energy reaches $\sim 10^{14}$ eV as shown in Figure \ref{fig:C},
which is consistent with the energy where cooling and
acceleration balances as $t'_{\rm c,syn}=1/K$.
The typical energy of synchrotron emission is in the gamma-ray range as
%%%%%%%%%%%%%%%%%%%%%%%
\begin{eqnarray}
\varepsilon_{\rm syn,typ}&=&\Gamma \frac{3 \hbar e B_0}{2 m_{\rm e} c} 
\gamma^{\prime 2}_{\rm e} \\
&\simeq& 7 \left( \frac{\Gamma}{100}\right)
\left( \frac{B_0}{0.1~\mbox{G}} \right)
\left( \frac{\varepsilon'_{\rm e}}{10^{14}~\mbox{eV}}\right)^2~\mbox{GeV}.
\end{eqnarray}
%%%%%%%%%%%%%%%%%%%%%%%
The gamma-ray photons {\it Fermi} detected are mainly emitted via synchrotron
emission in this model, while EIC emission by low-energy particles slightly contribute to
the gamma-ray flux.
The factor $\xi$ for those highest energy particles
is $\sim 3$ [see equations (\ref{eq:max}) and (\ref{eq:phmax})].

Below we discuss the details of the very complicated cascade process
with secondary pair injection,
re-acceleration, and the Klein--Nishina effect.
In model C, secondary electron--positron pairs produced
via $\gamma \gamma$-absorption are also accelerated by turbulence.
Even with a very low injection rate (Table \ref{table:para1}),
the secondary pair production and the re-acceleration
attain the electron energy density/distribution required to
reproduce the observed photon flux.
The short acceleration timescale boosts the maximum electron energy,
which leads to a high gamma-ray production rate at $\varepsilon'
>\varepsilon'_{\rm cut} \sim 100$ MeV,
where $\gamma \gamma$-absorption is efficient.
The number of electrons/positrons is largely dominated
by secondary pairs.
The synchrotron peak energy in the jet comoving frame ($\sim 70$ MeV)
is almost the same as the $\gamma \gamma$-cut-off energy
(see the left panel in Figure \ref{fig:fadd}).
Secondary electron--positron pairs are injected above $\varepsilon'_{\rm e}
\sim 70/2$ MeV$=35$ MeV.

The cooling time of electrons in the Klein--Nishina regime
is almost the same as the scattering timescale,
$t'_{\rm sc}=1/(n'_{\rm UV} \sigma_{\rm KN} c)$.
Adopting the photon density $n'_{\rm UV} \sim U'_{\rm UV}/(4 T'_{\rm UV})$
and the cross section $\sigma_{\rm KN} \sim (3/8) \sigma_{\rm T} x^{-1}
(\ln{2x}+1/2)$, where $x=4 T'_{\rm UV} \gamma'_{\rm e}/(m_{\rm e} c^2)
\simeq \gamma'_{\rm e}/130$,
we find that the synchrotron cooling becomes the dominant cooling process
for particles above $\varepsilon'_{\rm e} \sim 90$ GeV.

Around 70 MeV, the IC cooling timescale ($\sim 10^3$ s)
of electrons/positrons is comparable
to the acceleration timescale $K^{-1} \sim 2000$ s,
and the secondary injection is the most efficient in this energy range.
Above this energy, the IC cooling timescale grows with energy owing
to the Klein--Nishina effect.
Namely, the IC cooling timescale for electrons
between $\varepsilon'_{\rm e}=70$ MeV and 90 GeV
is longer than the acceleration timescale,
but still shorter than the synchrotron cooling timescale $t'_{\rm c,syn}
\propto \varepsilon^{\prime -1}_{\rm e}$.
In this electron energy range,
the radiative cooling, dominated by IC emission,
is the subdominant effect compared to the stochastic acceleration,
and the energy of IC photons emitted by such electrons is much
higher than the {\it Fermi} energy range.
For particles above $\varepsilon'_{\rm e}\sim 90$ GeV,
the energy loss rate due to synchrotron emission becomes larger
than the IC energy loss rate,
and the total cooling timescale starts to decrease with energy.
The cooling timescale
becomes comparable to the acceleration timescale again at $\varepsilon'_{\rm e}
\simeq 10^{14}$ eV.

If we neglect the re-acceleration and the Klein--Nishina effect
(the cooling timescale $\propto \gamma_{\rm e}^{-1}$),
the particle energy distribution becomes as soft as
$n'(\varepsilon'_{\rm e})\propto {\varepsilon}_{\rm e}^{\prime -(p+1)}$,
where $p \sim 2$ is the power-law index at the secondary injection.
The Klein--Nishina effect would make the spectrum harder than the above estimate.
On the other hand, if we only consider the acceleration and injection at
$\gamma'_{\rm inj}=10$,
the spectrum is proportional to ${\varepsilon}_{\rm e}^{\prime -1}$
in the steady hard-sphere model.
Our numerical result shows that
the complex combination of the above effects leads to a spectrum
slightly harder than ${\varepsilon}_{\rm e}^{\prime -2}$.

The synchrotron cooling time of the X-ray emitting electrons/positrons
($\varepsilon'_{\rm e}\simeq 3.9 \times 10^{10} (\varepsilon/1~\mbox{keV})^{1/2}
$ eV) is $t'_{\rm c,syn} \simeq 10^6 (\varepsilon/1~\mbox{keV})^{-1/2}$ s,
which is much longer than the dynamical timescale $t'_{\rm var}= R_0/(c \Gamma)
\simeq 2.4 \times 10^4$ s.
In this case, the dominant cooling process is
the adiabatic cooling, whose timescale is equal to the dynamical timescale.
Even in the X-ray band, the variability timescale is
regulated by the dynamical one $t_{\rm var} \simeq R_0/(c \Gamma^2)$
for an observer \citep[see][]{asa14}.
The synchrotron photon energy emitted by the
electrons whose cooling timescale satisfy $t'_{\rm c,syn}=t'_{\rm var}$
is $\sim 2$ MeV, which corresponds to the cooling break
in the photon spectrum in Figure \ref{fig:C}.

The synchrotron spectrum in model C reproduces
both the gamma-ray and X-ray data very well.
The gamma-ray spectrum is mainly produced by synchrotron emission from
the secondary pairs
with a partial contribution of EIC emission.
If we neglect the secondary pairs, the gamma-ray spectrum
becomes dim and hard as shown by the blue line in
the right panel of Figure \ref{fig:C}.
This synchrotron model with electromagnetic cascade is a very unique model
to account for bright gamma-ray emission.
To emit gamma rays via synchrotron by electrons, a strong magnetic field
is not necessarily required.

Recently, \citet{abd19} reported the sub-TeV gamma-ray detection
by H.E.S.S. from the same 2015 June flare, though the observation time
was not simultaneous with orbits C and D (about 13 hr later).
In our model C, the highest energy of electrons is 100 TeV,
and most of the very-high-energy gamma-ray photons with energies $\varepsilon \gg 7$ GeV
generated via EIC emission are absorbed in the source.
A small fraction of such photons escape from the source
as seen in Figure \ref{fig:C}.
The sharp cut-off above 100 GeV is due to EBL absorption.
The sub-TeV photon flux in our model seems consistent with
the observed flux, a few times $10^{-11}~\mbox{erg}~\mbox{cm}^{-2}~\mbox{s}^{-1}$.

\section{Summary and Discussion}
\label{sec:sum}

The very bright gamma-ray flare of 3C 279 in 2015 June
can be explained by synchrotron emission.
The assumed parameters $R_0=7.1\times 10^{16}$ cm
and $\Gamma=100$ are consistent with the variability timescale.
Motivated by the magnetic reconnection model,
we have tested prompt electron injection with a power-law energy distribution
with exponential cut-off.
The required maximum energy of electrons is close to the limit of $\xi=1$.
If we adopt a strong magnetic field, the synchrotron emission
from secondary electron--positron pairs is unavoidable.
In order to reconcile the X-ray and UV fluxes, the magnetic field
is at most 8 G.
The production efficiency of secondary particles depends on
the detail of the high-energy cut-off shape of the injected electron spectrum
\citep{aha86,zir07}.
However, to reduce the secondary synchrotron flux by adjusting the cut-off shape,
a fine tuned parameter set is required.
A largely Poynting-flux-dominated jet
is unlikely in terms of the X-ray spectrum and the energy budget.

We have also considered the stochastic acceleration model,
in which the particle acceleration is phenomenologically
expressed by the diffusion coefficient $D_{\gamma \gamma}$.
Thanks to our time-dependent code,
we have obtained a very unique picture
accompanying the electromagnetic cascade and re-acceleration
of the secondary electron--positron pairs.
The magnetization in this model is very low as $U_B/U_{\rm e}\sim 10^{-4}$.
Therefore, the synchrotron model for this flare does not necessarily mean
a higher magnetization than the typical values in other blazars.

In our stochastic acceleration model with the electromagnetic cascade,
the acceleration timescale, $K^{-1} \sim 2000$ s,
does not depend on the electron energy.
This type of acceleration 
can be realized by wave-particle interaction via transit-time damping.
Considering the negligible energy fraction of the magnetic energy,
let us focus on the fast MHD waves, in which the wave energy
is dominated by kinetic energy rather than magnetic energy.
If the turbulence injected at the scale of $R_0/\Gamma$
with the relativistic sound velocity $c/\sqrt{3}$
cascades to shorter scales with a shortest scale $\lambda_{\rm min}$,
according to the simulation in \citet{ter19}, the diffusion coefficient is
%%%%%%%%%%%%%%%%%%%%%%%
\begin{eqnarray}
K \simeq 5.6 \frac{\pi}{18} c \left(\frac{R_0}{\Gamma} \right)^{-1/3} \lambda_{\rm min}^{-2/3},
\end{eqnarray}
%%%%%%%%%%%%%%%%%%%%%%%
where the Kolmogorov turbulence is assumed.
Our model C requires $\lambda_{\rm min}=2.2 \times 10^{13}$ cm,
which is 3\% of the injected scale $R_0/\Gamma$.
The shortest wave length $\lambda_{\rm min}$ should be longer than the Larmor radius of electrons
to realize the hard-sphere acceleration.
From $B_0=0.1$ G and the highest electron energy $10^{14}$ eV,
we obtain a Larmor radius $3.3 \times 10^{12}$ cm,
which is consistently shorter than $\lambda_{\rm min}$.
As a matter of course, the implied value of the acceleration timescale parameter
is $\xi \sim 3$ at $10^{14}$ eV as we have mentioned in the previous section.
Hydrodynamical eddy turbulences with a similar scale also
induce the hard-sphere acceleration.
Although we have an unknown parameter $\lambda_{\rm min}$,
which may depend on the detail of the time-dependent energy transfer process
between waves and particles with back-reaction,
the required acceleration efficiency seems consistent with
the turbulence acceleration picture.

\begin{acknowledgments}
%We also thank F. Takahara, M. Kusunose, K. Toma,
%J. Kakuwa, K. Nalewajko and G. M. Madejski for useful discussion.
The authors thank the anonymous referees for the valuable and helpful comments.
This work is supported by %JSPS KAKENHI grant No. 18K03665,
the joint research program
of the Institute for Cosmic Ray Research (ICRR),
the University of Tokyo.
\end{acknowledgments}

\end{document}